# Rotational and translational waves in a bowed string

*Eric Bavu, Manfred Yew, Pierre-Yves Plaçais, John Smith and Joe Wolfe*

School of Physics, University of New South Wales, Sydney Australia
J.Wolfe@unsw.edu.au

## Abstract

We measure and compare the rotational and transverse velocity of a bowed string. When bowed by an experienced player, the torsional motion is phase-locked to the transverse waves, producing highly periodic motion. The spectrum of the torsional motion includes the fundamental and harmonics of the transverse wave, with strong formants at the natural frequencies of the torsional standing waves in the whole string. Volunteers with no experience on bowed string instruments, however, often produced non-periodic motion. We present sound files of both the transverse and torsional velocity signals of well-bowed strings. The torsional signal has not only the pitch of the transverse signal, but it sounds recognisably like a bowed string, probably because of its rich harmonic structure and the transients and amplitude envelope produced by bowing.

## 1. Introduction

Torsional waves are excited in strings because the friction with the bow acts at the surface of the string, rather than at its centre of mass. The *n*th harmonic of a torsional standing wave has considerably higher frequency than that of a transverse standing wave, and there is in general no harmonic relation between waves of the two types. Torsional waves exert little torque on the bridge and so have little direct acoustic effect. However, the velocity of the bow-string contact is a linear combination of transverse and angular velocities, and this velocity determines the transitions between stick and slip phases. Consequently, in some bowing regimes, torsional waves may produce non-periodic motion or jitter. Because of the ear is very sensitive to jitter, this can have a large effect on the perceived sound.

A number of references describe aspects of the physics of the bowed string [1-8]. Helmholtz motion is the idealised, steady state motion of a one-dimensional string between two completely fixed boundaries, excited by a bow with negligible width (Fig 1).

The transverse wave speed is $\sqrt{F/\mu}$, where $F$ is the tension and $\mu$ the line density, so the frequency of standing waves is adjusted by changing $F$, e.g. when one tunes the string. The string can also support torsional standing waves, whose speed is $\sqrt{K/I}$, where $K$ is the torsional stiffness and $I$ the specific moment of inertia. The frequencies of torsional standing waves are only very weakly dependent on tension. In general, therefore, there is no harmonic relationship between the frequencies of transverse and torsional waves. The relative motion of the bow and string, and therefore the times of commencement of stick and slip phases, depends on both the transverse and torsional velocity (Fig 2). Consequently, torsional waves can introduce aperiodicity or jitter to the motion of the string. Human hearing is very sensitive to jitter [9]. Small amounts of jitter contribute to a sound's being identified as 'natural' rather than 'mechanical'. Large amounts of jitter, on the other hand, sound unmusical. Here we measure and compare translational and torsional velocities of a bowed bass string and examine the periodicity of the standing waves.

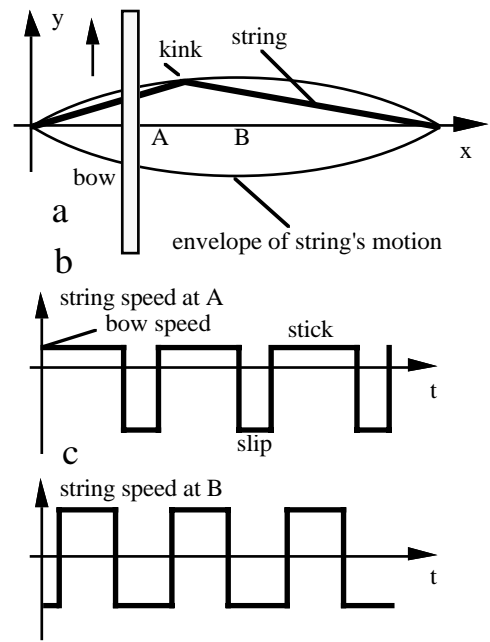

*Figure 1*: The (translational) Helmholtz motion of an idealised, one-dimensional string. The shape of the string is always two straight lines, which are joined by a kink that travels around the envelope shown (a). At the bowing point (A), an upwards moving bow starts the stick phase at $t = 0$ with $v > 0$ (b), during which the kink travels to the right hand end of the string, reflects and returns. It returns to A, and the large transient force starts the slip phase ($v < 0$) during which the kink travels to the left hand end, is reflected and returns to the bow, ready to begin the stick phase again. At the midpoint of the string (B), the up and down speeds are equal (c). See the animation [10].

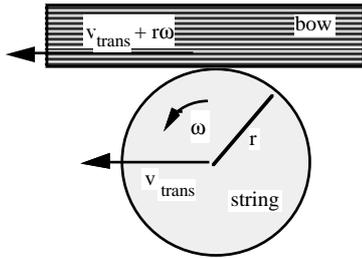

*Figure 2. A string of radius r. During the stick phase, the bow speed is approximately $v_{trans} + r\omega$.*

## 2. Materials and methods

A permanent magnet was fitted with pole pieces shaped to give a localised region with homogeneous field [11]. The field in the region of the string was mapped with a Hall probe. The region with inhomogeneity of less than 1% was a circle of diameter 14 mm in the plane of vibration of the string. A bass guitar string (low E or E1), with steel windings on a steel core, was strung over two rigid 'bridges' separated by $L$ = 750 mm, so that it passed through the centre of this region: hereafter the measurement point. Two sections of fine, insulated copper wire, 12 mm long, were attached to the string longitudinally on opposite sides, near the top and bottom of the string, with mylar tape. Their ends were connected to a measurement circuit via semicircular loops whose plane had zero magnetic flux. The emfs of these wires are proportional to $v_{trans} \pm r\omega$. They were measured with a digital oscilloscope, whence the data were transferred to a computer. The average emf is used as a measure of the transverse velocity of the centre of the string. The constant of proportionality was determined by oscillating the measurement circuit sinusoidally with an accelerometer attached to a uniaxial shaker. In principle, the difference between these two emfs gives $r\omega$, but the signal:noise ratio is inadequate. Copper wire, 0.05 mm diameter, was wound in a coil 10 turns, 12 mm long and 2 mm wide, and attached to the surface of the string, oriented with its normal perpendicular to both the string and the field [12]. The flux linkage of this coil was calibrated using an oscillating magnetic field and a Hall probe.

The position of the measurement point is a compromise: close to the bridge produces small amplitude waves and poor signal:noise ratio. Too far from the bridge, however, and the measurement point approaches the nodes of low harmonics and therefore loses high frequency information. For this study, the measurement point was at set at $L/5$. The string was bowed with a violin bow (chosen over a bass bow because of its narrower width) at points displaced from the bridge a distance $L/m$, where $m$ = 6, 7... 15. ($L/6$ is as close as practical from the centre of the coil). For the results shown here, the bow was operated by a violinist (EB) with 17 years experience.

The frequency of free torsional standing waves was determined by twisting the string near the measurement position and then releasing it, and examining the oscillogram and Fourier transform of oscillations measured by the coil.

## 3. Results and conclusions

A range of different tensions was used. The tension chosen for most of the studies gave a fundamental frequency for the translational motion of 40 Hz. The fundamental frequency of a free torsional wave on the whole string length was largely independent of tension, and had the value 225 Hz, and so a little below the sixth harmonic of the translational fundamental. The finite length ($d$ = 12 mm) of the wires and coils used to measure translation and rotation limit the measurable minimum wavelength to a few times d. We found that the translational signals measured for the harmonics above about the tenth are strongly attenuated, which may be a limitation of this measurement technique. Fortunately, much interesting behaviour is observable within this constraint.

Fig 3 shows a typical result. The string was bowed by an expert string player at $L/13$ and measured at $L/5$. The translational wave is highly periodic and globally exhibits the characteristics of Helmholtz motion: a long, slow movement in one direction ($v > 0$ here) and a short, fast return. The oscillations in the measurement during each of these phases are in part the result of the low pass filtering caused by the finite length of the sensor: a rectangular wave with the high harmonics missing looks just like a rectangular wave with spurious high harmonics (with opposite phase) added. The analogous limitation in the torsional signal occurs at a much higher frequency: the tenth harmonic of the torsional fundamental is above 2 kHz.

The spectrum of the translational wave looks approximately like that for low-pass filtered Helmholtz motion measured at $L/5$, except that the fifth, and tenth harmonics are not quite zero.

The torsional wave has strong, sharp oscillations near the times of the change in direction of transverse motion: this is true of all the measurements made. The maximum amplitude of $r\omega$ is about 20 times smaller than the v of the fast translation phase, or about 6 times smaller than the $v$ of the slow phase. The torsional wave is approximately periodic over the same period (25 ms) as the transverse wave.

This is clearer in the frequency domain, where the torsional wave is seen to have nearly all of the harmonics of the *transverse* fundamental. The torsional resonances of the string are evident as formants in this spectrum: there is a strong sixth harmonic (the fundamental of the free torsional wave) and strong 11th, 12th and 13th harmonics (around the second harmonic of free rotation). In experiments in which the tension was changed, the formants remained at similar frequencies while the pitch varied as expected; consequently, the formants appeared on different harmonics of the translational fundamental.

Fletcher [13] shows that phase locking of two inharmonic oscillators is most readily achieved (i) at low harmonics, (ii) when the frequencies are in close to integer ratios and (iii) when the coupling interaction is strongly nonlinear. Here: (i) the transverse fundamental locks the motion, (ii) the ratio lies between 5 and 7 and (iii) the friction is very strongly nonlinear. So mode locking is not surprising.

In all our results, when the string was bowed at L/m, the torsional spectrum always had a strong peak at a frequency equal to that of the $m$th harmonic of the translational fundamental. Further, while Helmholtz [1] observed that a string bowed at $L/m$ produced a transverse motion from whose spectrum the $m$th harmonic and all its integral multiples were missing, this was not observed in our results.

Another way of presenting the results is as sound files [10]. In comparing with acoustic instruments, one should remember that these are velocity signals, and so are filtered at 6 dB/octave with respect to the signal of the force at the bridge, and that they don't include the complicated radiativity of the instrument. (In the sensitive range of hearing, the radiativity of a violin decreases at ~ 9dB/octave [14].) As one would expect for periodic signals, they have a clear pitch. The sound of the translational velocity is not unlike the sound of an electrified, bowed bass (The signals from electric coil pickups are approximately proportional to velocity). The sound of the rotational velocity has, of course, much less bass, but it is recognisably the sound of a bowed string instrument, probably because it is harmonically rich and it shares the transients and envelope of the transverse signal. It has the same pitch as the translational velocity, but the formants are so strong that at least one may often be identified in the sound.

We noted above that $\omega$ has a peak near the end of the slow part of the transverse motion at which $r\omega$ is not negligible in comparison with the transverse speed. For the bowing point, this corresponds to the end of the stick phase. To the extent that the torsional wave is *not* periodic at the period of the transverse wave, one would expect variations in $r\omega$ from one cycle to the next to introduce jitter.

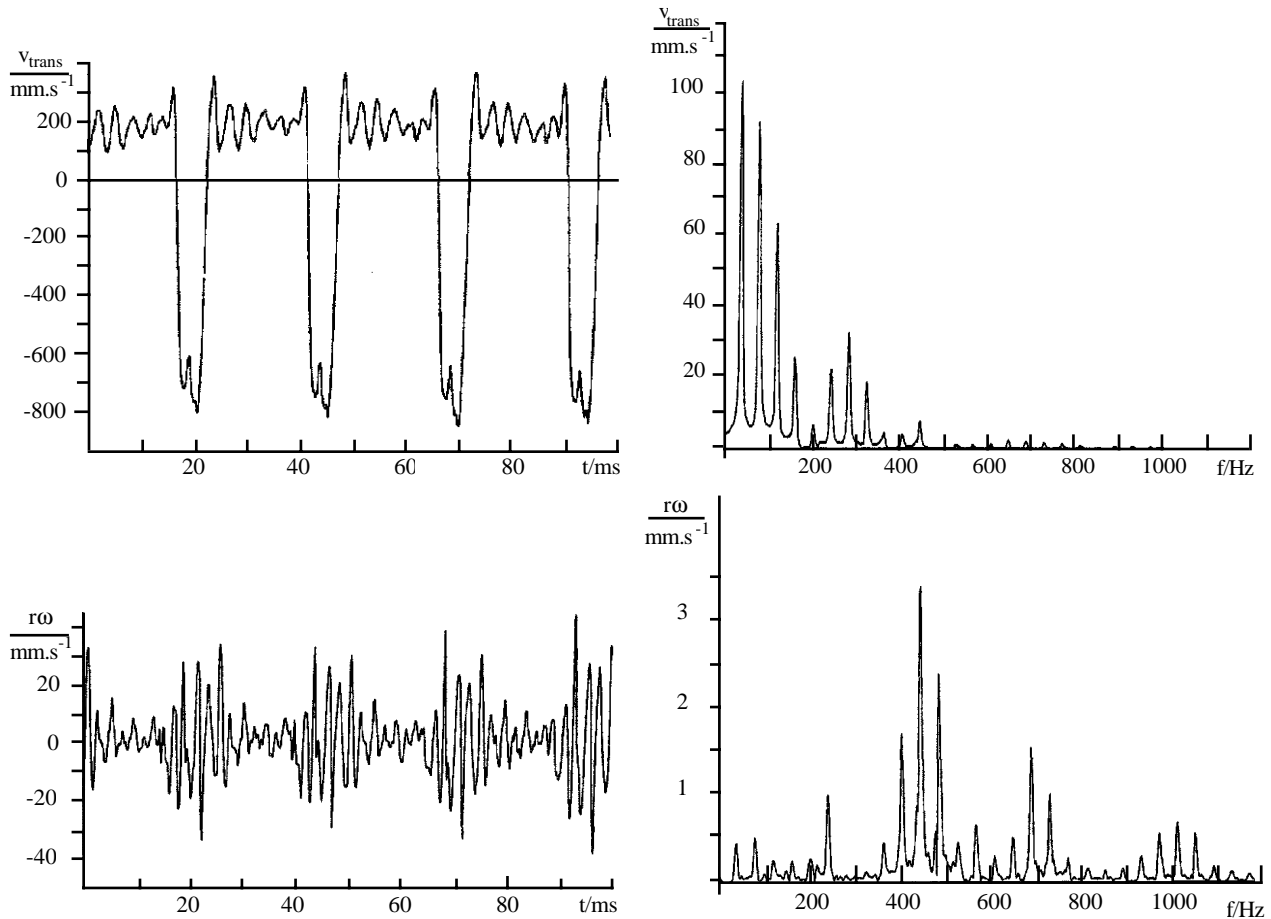

*Figure 3*: T*ime domain (left) and frequency domain (right) representations of the signals measured for v and rω for a string bowed at L/13 and measured at L/5. The translational signal is inherently low-pass filtered in this technique.*

Because of the ear's sensitivity to jitter, this is potentially an important acoustic effect. However, the transverse wave is very nearly periodic (and so jitter is approximately zero). This periodicity, along with the period and formants in the torsional signal, are consistent with phase locking of the torsional motion to the (stronger) translational motion. The non-linearity required for phase locking is due here to the frictional coupling between bow and string.

The requirement of phase locking may contribute to the constraints on the relation between bow force and speed [5]. Experiments with inexperienced players often produced non-periodic motion. One possible explanation is that they were incapable of producing the parameters necessary for phase locking.

## 4. Conclusions

The torsional waves in the bowed bass string contribute a component to the surface speed that, at typical bowing positions, is not negligible in comparison with the (translational) Helmholtz motion. However, torsional waves do not introduce much jitter because, in a skilfully bowed string, they are phase locked to the translational waves.

## Acknowledgments

This research includes undergraduate research projects of EB, MY and P-YP.

## Appendix: Torsional Helmholtz motion

We show above that, in a well bowed string, $r\omega$ phase locks to the larger signal $v_{trans}$. This invites one to wonder whether, if the amplitude inequality were reversed, $v_{trans}$ might phase lock to $r\omega$. This condition was investigated by constraining a string at the bowing point with a thin teflon block that inhibited waves near the bow.

Fig 4 shows Helmholtz motion in the torsional wave, at the fundamental frequency of the unconstrained torsional wave. $v_{trans}$ is rather smaller than $r\omega$, and $v_{trans}$ is indeed phase locked to $r\omega$. The musical usefulness of this phenomenon is limited, but it is an interesting example of the mode locking described by Fletcher[13].

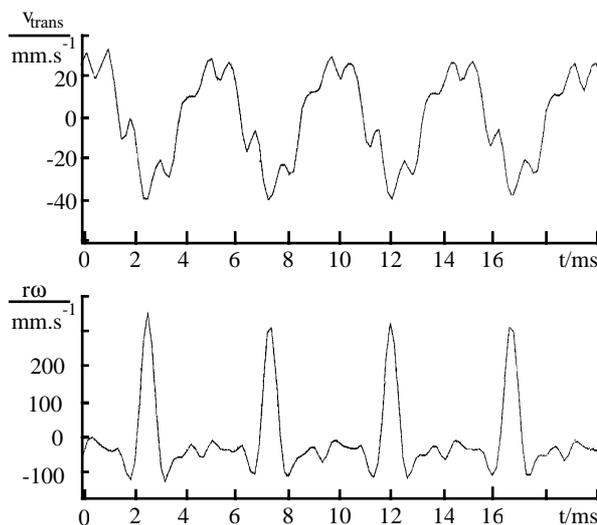

*Figure 4. Helmholtz motion in torsional waves can phase lock transverse motion. Note the scale changes compared with Fig 3.*